\documentclass{ws-procs961x669}            
\begin{document}
\title{Effective field theory from Relativistic Generalized Uncertainty }

\author{Vasil N. Todorinov$^*$, Saurya Das, and Pasquale Bosso}

\address{Department of Physics \& astronomy, University of Lethbridge,\\
Lethbridge, AB T1K 3M4, Canada\\
$^*$E-mail: v.todorinov@uleth.ca\\
https://www.ulethbridge.ca}

\begin{abstract}
Theories of Quantum Gravity predict a minimum measurable length and a corresponding modification of the Heisenberg Uncertainty Principle to the so-called Generalized Uncertainty Principle (GUP). However, this modification is usually formulated in non-relativistic language, making it unclear whether the minimum length is Lorentz invariant. We have formulated a Relativistic Generalized Uncertainty Principle, resulting in a Lorentz invariant minimum measurable length and the resolution of the composition law problem. This proved to be an important step in the formulation of Quantum Field Theory with minimum length. We derived the Lagrangians consistent with the existence of minimal length and describing the behaviour of scalar, spinor, and U(1) gauge fields. We calculated the Feynman rules (propagators and vertices) associated with these Lagrangians. Furthermore, we calculated the Quantum Gravity corrected scattering cross-sections for a lepton-lepton scattering. Finally, we compared our results with current experiments, which allowed us to improve the bounds on the scale at which quantum gravity phenomena will become relevant.
\end{abstract}


\bodymatter

\section{Introduction}\label{sec:Introduction}
The quantization of gravity is one of the major outstanding problems of fundamental physics. The family of theories tackling this problem is known as Quantum Gravity (QG). This family is home to a large diversity of ideas and approaches to the problem. The most prevalent members of the family, namely String Theory, Loop Quantum Gravity in addition to models such as  Doubly-Special Relativity theories as well as many others, agree on the existence of  a minimum measurable length or a similar scale in spacetime. The minimum length is usually considered to be the proportional to the
Planck length, $\ell_{Pl}=10^{-35}m$, and it signifies the scale at which quantum gravity effects would manifest. Currently, we have no access to energy regimes anywhere close to the Planck scale, therefore all theories of QG lack a direct experimental confirmation. The field of Quantum Gravity Phenomenology (QGP) tackles this problem by using the existing theories to make models in order of finding low energy remnants of QG phenomena in already existing experiments. 

As we already mentioned, one phenomenon that most theories of QG agree on is the existence of a minimum measurable length, which is in direct contradiction with the Heisenberg uncertainty principle (HUP), since the latter allows for infinitely small uncertainties in position. The solution is the introduction of a new model which deforms the HUP in order to accommodate for a minimal uncertainty in position. The deformed HUP is known as Generalized Uncertainty Principle (GUP), and was first introduced by Kempf, Mangano, and Mann in 1995 \cite{Kempf:1994su}. One of the possible position-momentum commutators corresponding to a GUP model has the following form
\begin{equation}\label{eq:KMM}
         [x_i,p_j]=i\hbar\delta_{ij}(1+\beta_1 p^2)+i\hbar\beta_2 p_i p_j\,,
\end{equation}
where $\beta_1=\beta_1^*(\ell_\text{Pl}/\hbar)^2$, $\beta_2=\beta_2^*(\ell_\text{Pl}/\hbar)^2$, and $\beta_1^*,\beta_2^*$ are numerical coefficients. The GUP model has been successful in modeling a wide array of corrections to quantum mechanical phenomena \cite{Adler1999-db,Adler_2001,Ali2011,Ali_2014,Ali_2015,Alonso_Serrano_2018,Amati1989-gs,Amelino_Camelia_2001,AMELINO_CAMELIA_2002,Amelino-Camelia2013-xs,Bargue_o_2015,Bambi2007-te,Bawaj_2015,Bojowald2011-bb,Bolen2005-jq,bosso2017generalized,Bosso2018,Bosso:2018uus,Bosso:2019ljf,Burger_2018,bushev2019testing,Capozziello:1999wx,Casadio_2020,Chang:2011jj,Cortes:2004qn,Costa_Filho2016-ox,Dabrowski2019-cb,Dabrowski2020-kk,Das2008,Das:2009hs,Das:2010zf,DAS2011596,Das2014,Das_2019,Das_2020,Garcia-Chung:2020zyq,Giddings2020-xz,Hamil2019-qh,Hossenfelder:2006cw,Hossenfelder:2012jw,Kempf:1994su,Kober:2010sj,KONISHI1990276,MAGGIORE199365,Maggiore:1994,Marin:2013pga,Mead1966-xj,Moradpour2021-jy,Mureika2019-lf,Myung_2007,Park2008-uj,Scardigli1999-ne,Snyder:1946qz,Sprenger_2011,Sriramkumar:2006qt,Stargen_2019,tedesco2011fine,wang2016solutions}, and has provided compelling bound on the minimum length scale, by comparing the result to existing experimental data.

However, the model introduced by Eq. \eqref{eq:KMM} has two shortcomings:
\begin{itemize}
    \item The position-momentum commutator and the corresponding uncertainty relation are not Lorentz covariant, and therefore the resulting minimum length is frame dependent,
    \item \textit{Composition Law Problem}: Energies and momenta do not sum up linearly \cite{Hossenfelder:2014ifa}.
\end{itemize}

The composition law problem arises due to the fact that  deformation of the HUP is done through a generalization of the momentum operator. Namely terms of higher order in momentum are added in order to deform the commutator. This  leads to a deformation of the Einstein dispersion relation of the form
\begin{equation}
E^2=p^2 c^2+m^2 c^4 ~\rightarrow~E^2=p^2c^2+m^2c^4+\mathcal{O}(p^4) \,.
\end{equation}
Consequently one can show that if there are terms of higher than second order in momentum in the dispersion relation, for a composite system moving in the same speed the energies and momenta do not sum up linearly,
\begin{equation}
    \vec{p}_3\neq  \vec{p}_2+\vec{p}_1~~~~~ E_3\neq E_1+E_2\,.
\end{equation} 
It is important to note that solutions of this problem have been developed, some examples are \cite{Amelino-Camelia:2014gga,Lake:2019oaz}.

Such problems prevents us to apply the GUP models to Relativistic Quantum Mechanics and Quantum Field Theory, which essentially restricts the energy regimes in which we can search for the  remnants of QG effects, by preventing us from using the highest energy experiments currently available, \textit{i.e.} LHC. 

In the following sections, we will address these problems by extending the GUP up to second order in Planck length, to relativistic regimes (RGUP), we will explore the deformations of the  Lorenz group and symmetries associated with it, and we will obtain Lagrangians allowing for the existence of minimum length for fields of spin-$0,1/2,1$.
\section{Relativistic Generalized Uncertainty Principle\cite{Todorinov2018-xi}}\label{sec:RGUP}

Inspired by \cite{Quesne2006}, we will use the most general form of quadratic GUP algebra and extend it to a covariant expression in Minkowski spacetime with the following signature  $\{-,+,+,+\}$.
The particular form obtained for the commutator is as follows
\begin{equation}
\label{GUPxp}[x^{\mu},p^{\nu}]=i\hbar\,\left(1+(\varepsilon-\kappa)\gamma p^{\rho}p_{\rho}\right)\eta^{\mu\nu}+i\,\hbar(\beta+2\varepsilon)\gamma p^{\mu}p^{\nu}\,,\\
\end{equation}
where $\mu\,,\nu\in\{0,1,2,3\}.$ For convenience it is useful to separate the dimensionality of the corrections into a dimensionful parameter $\gamma$, with the dimension of inverse momentum, and dimensionless parameters $\kappa,\beta$ and $\epsilon$ (and also $\xi$ in Eq.(\ref{GUPx}) below).
The scale of the QG effects $\gamma$ is defined as $\gamma=\frac{1}{(c\,M_{Pl})^2}=\frac{\ell_{\text{Pl}}^2}{\hbar^2}$, where $M_{Pl}$ is the Planck mass, and $\ell_{\text{Pl}}$ is the Planck length.
The minimum length we can obtain from the algebra Eq.(\ref{GUPxp}) 
is of the order of the Planck length.
Important to note is that Eq.\eqref{GUPxp} reduces in the non-relativistic ($c\rightarrow \infty$) limit to the one proposed in 
\cite{Kempf:1994su}. Furthermore, in the non-relativistic ($c\rightarrow \infty$), and non-GUP limit 
($\gamma \rightarrow 0$) limit, Eq.\eqref{GUPxp} reduces to the standard Heisenberg  algebra. 
We can clearly see from the commutator Eq.\eqref{GUPxp} that while $x^{\mu}$ and $p^{\nu}$ are the physical position and momentum, they are no longer canonically conjugate variables.
Therefore, we introduce 
two auxiliary $4$-vectors, 
$x_0^\mu$ and $p_0^\nu$ which are 
canonically conjugate, such that 
\begin{align}
\label{MathVar}p_0^{\mu} = & -i\hbar\frac{\partial}{\partial x_{0\,\mu}}, &
[x_0^{\mu},p_0^{\nu}] = & i\hbar\eta^{\mu\nu}.
\end{align}\\
Then we assume that the physical position and momentum are both functions of the auxiliary ones,
\begin{equation}\label{Eq:PositionMomentumFunctions}
 x^\mu=x^\mu(x_0,p_0),\quad
 p^\mu=p^\mu(x_0,p_0)\,.
 \end{equation}
By expanding Eq. \eqref{Eq:PositionMomentumFunctions}  in Taylor series, which we truncate up to first order in the QG scale $\gamma$, and imposition of commutativity in momentum space $[p_\mu,p_\nu]=0$, we can write the general expressions of the physical position and momentum are as follows
\begin{align}
\label{GUPx}x^{\mu}&=x_0^{\mu}-\kappa\gamma p_0^{\rho}p_{0\rho}x_0^{\mu}+\beta\gamma p_0^{\mu}p_0^{\rho}x_{0\rho}+\xi\hbar\gamma p_0^{\mu}, \\
\label{GUPp}p^{\mu}&=p_0^\mu\,(1+\varepsilon\gamma p_0^{\rho}p_{0\rho})\,.
\end{align}
\\
Using Eqs.(\ref{GUPx}) and (\ref{GUPp}), 
we can calculate the following expression for the position-position commutator 
\begin{equation}
    \label{GUPxx}[x^{\mu},x^{\nu}]=
i\hbar\gamma\frac{-2\kappa+\beta}{1+(\varepsilon-\kappa)\gamma p^{\rho}p_{\rho}}\left(x^{\mu}p^{\nu}-x^{\nu}p^{\mu}\right).
\end{equation}
Therefore as in \cite{Kempf:1994su}, 
in this case  as well we arrive at a non-commutative spacetime.

We also note that the last two terms in Eq.\eqref{GUPx}  introduce a preferred direction of $p_0^\mu$ and therefore break isotropy of spacetime.
Since this violates the principles of relativity, 
we will assume that $\beta=\xi=0$ from now on.

\subsection{Lorentz and Poincar\'e algebra}\label{sec:LnP} 
 Using Eqs.\eqref{GUPx} and \eqref{GUPp}, we now construct the generators of the Lorentz group
\begin{equation}
    M^{\mu\nu} = p^{\mu}x^{\nu}-p^{\nu}x^{\mu}
    = \left[1+(\varepsilon-\kappa)\gamma p_0^{\rho}p_{0\,\rho}\right]\tilde{M}^{\mu\nu}\,,
\end{equation}
where $\tilde{M}^{\mu\nu}=p_0^{\mu}x_0^{\nu}-p_0^{\nu}x_0^{\mu}$ is the Lorentz generators constructed from the canonical variables  $x_0$  and $p_0$. The commutator representative of the Poincar\'e algebra for the physical operators can then be calculated
\begin{subequations}\label{eq:Poincare}
\begin{align}
 [x^\mu,M^{\nu\rho}] &=  i\hbar[1 + (\varepsilon - \kappa) \gamma p^{\rho} p_{\,\rho}]\left(x^{\nu}\delta^{\mu\rho}-x^{\rho}\delta^{\mu\nu}\right) + i\hbar 2 (\varepsilon - \kappa) \gamma p^{\mu} M^{\nu\rho}\label{xM}\\
   \label{pM}[p^\mu, M^{\nu\rho}]& =  i\hbar[1 + (\varepsilon - \kappa) \gamma p^{\rho} p_{\,\rho}]\left(p^{\nu}\delta^{\mu\rho}-p^{\rho}\delta^{\mu\nu}\right),
    \\ \label{MM} [M^{\mu\nu},M^{\rho\sigma}] &= i\hbar\left(1 + (\varepsilon - \kappa) \gamma p^{\rho} p_{\,\rho}\right)\left(\eta^{\mu\rho}M^{\nu\sigma}
    -\eta^{\mu\sigma} M^{\nu\rho}-\eta^{\nu\rho}M^{\mu\sigma}+\eta^{\nu\sigma}M^{\mu\rho}\right)\,.
\end{align}
\end{subequations}
From 
Eqs.(\ref{xM},\ref{pM},\ref{MM}),
we can see that on the line 
$\varepsilon=\kappa$ in the parameter space, one has a non-trivial RGUP with an {\it unmodified} Poincar\'e algebra. 
On this line in parameter space, the RGUP algebra
and the non-commutativity of spacetime coming from 
Eqs.(\ref{GUPxp}) and (\ref{GUPxx}) takes this form:
\begin{align}
[x^{\mu},p^{\nu}]&=i\hbar\,\left(\eta^{\mu\nu}+2\kappa\gamma p^{\mu}p^{\nu}\right)\,, \label{up1} \\
[x^{\mu},x^{\nu}]&=-
2i\hbar\kappa\gamma\left(x^{\mu}p^{\nu}-x^{\nu}p^{\mu}\right)\,, \label{nc1}
\end{align}
where $\kappa>0$.
We notice that for any one-dimensional spatial component, Eq.(\ref{up1}) mirrors the one obtained in  \cite{Kempf:1994su}.
Eqs.(\ref{up1}) and (\ref{nc1})
have similarities to the ones proposed 
in \cite{Snyder:1946qz}.
\subsection{Lorentz invariant minimum length and the Composition Law Problem}\label{sec:Poblems} 
The main reason for extending GUP to relativistic frameworks is to make its minimum uncertainty in position Lorentz invariant, or in other words to make a minimum length frame independent.  
In relativistic QM, position and momentum are promoted to operators and the Lorentz transformations $\Lambda$ are represented by a unitary operator $U(p^\nu,M^{\rho\sigma})$, {\it i.e.} $U^*=U^{-1}$, where $p^\nu$ is the generator of translation and $M^{\rho\sigma}$
is the generator of rotations of the Poincar\'e group. The position and momentum in this case transform as follows 
\begin{align}
    x^{\prime\mu}&=U(p^\nu,M^{\rho\sigma})x^\mu U^{-1}(p^\nu,M^{\rho\sigma})\,,\\
     p^{\prime\mu}&=U(p^\nu,M^{\rho\sigma})p^\mu U^{-1}(p^\nu,M^{\rho\sigma})\,.
\end{align}
The operator $U(p^\nu,M^{\rho\sigma})$ can be expressed from the generators of the Poincar\'e algebra as
\begin{equation}
    U(p^\nu,M^{\rho\sigma})=\exp\left[i a_\nu p^\nu\right]\exp\left[i \frac{\omega_{\rho\sigma}M^{\rho\sigma}}{2}\right]\,,
\end{equation}
where the coordinate system is translated by a vector $a_\mu$ and is rotated by $\omega_{\rho\sigma}$. 

The position-momentum commutator transforms as follows
\begin{equation}\label{Eq:TransformedComutator}
    [x^{\prime\mu},p^{\prime\nu}]=U[x^{\mu},p^{\nu}] U^{-1}\,.
\end{equation}
Substituting $[x^{\mu},p^{\nu}]$ for its expression as given in Eq.\eqref{GUPxp}, we get 
\begin{subequations}
\begin{align}
   [x^{\prime\mu},p^{\prime\nu}]&=U \left\{i\hbar\,\left(1+(\varepsilon-\kappa)\gamma p^{\rho}p_{\rho}\right)\eta^{\mu\nu}+i\,\hbar(\beta+2\varepsilon)\gamma p^{\mu}p^{\nu}\right\} U^{-1}\\
    &=i\hbar\,\left(1+(\varepsilon-\kappa)\gamma p^{\prime\rho}p^{\prime}_{\rho}\right)\eta^{\mu\nu}+i\,\hbar(\beta+2\varepsilon)\gamma U p^{\mu} U^{-1}Up^{\nu} U^{-1}\\
      &=i\hbar\,\left(1+(\varepsilon-\kappa)\gamma p^{\prime\rho}p^{\prime}_{\rho}\right)\eta^{\mu\nu}+i\,\hbar(\beta+2\varepsilon)\gamma p^{\prime\mu} p^{\prime\nu} \,.
\end{align}
\end{subequations}
Therefore, the commutator between position and momentum has the same form in every frame. Then, we can safely conclude that every frame will observe the same minimum measurable length.

For the composition law problem, we need to find the dispersion relation and see if it is quadratic or it has terms of higher order in momentum. As we know, the dispersion relation is connected to the Casimir invariants of the Poincar\'e algebra. In the unmodified case, there are two Casimir invariants: the momentum squared $p_0^\mu p_{0\mu}$; and the physical Pauli-Lubanski vector squared $\tilde{W}^\mu \tilde{W}_\mu$, where
\begin{equation}\tilde{W}_\mu\equiv \frac{1}{2}\varepsilon_{\mu \nu \rho \sigma }\tilde{M}^{{\nu \rho }}p_0^{\sigma}\,.\end{equation}
In our case, both the Poincare algebra and the momentum and Pauli-Lubanski vectors are deformed. However, by taking the commutator between the physical momentum squared $ p^\mu p_{\mu}$ and all the other generators of the algebra, it can easily be shown that it is a Casimir invariant of the modified Poincar\'e algebra presented in Eq.\eqref{eq:Poincare}. This can be repeated for the Pauli-Lubanski vector squared as well. 

This shows that our dispersion relation can be written as 
\begin{equation}
    p_\mu p^\mu=-m^2c^4\,.
\end{equation}
As we can see, the dispersion relation is quadratic in the physical momentum which means that energies and momenta sum up linearly as they should. It is important to note that due to the fact that the physical momentum is a polynomial of the auxiliary one, the resulting equations of motion will be of higher than second order. 
\section{Effective Quantum Field Theory with minimum length\cite{Bosso2020-dv,Bosso2020-fz}}\label{sec:EQFTwML} 
Having cleared up the frame dependence and the composition law problem for our model, we can proceed to apply it to quantum field theory. 

In the general case, the commutation relation we get for the position-position operator depends on the momentum. Therefore, it does not form an algebra and it cannot be represented by a smooth Lie manifold. For that reason we have chosen for further calculations to fix the model by making $\kappa=0$ and $\epsilon=\gamma_0$. We can redefine the RGUP parameter as follows
 \begin{equation}
    \gamma=\frac{\gamma_0}{(M_{\text{Pl}}c)^2}\,.
\end{equation}
In this particular model, the physical position and momentum take the form
 \begin{subequations}
\begin{align}\label{Eq:XPParticualrCase}
x^{\mu}&=x_0^{\mu}\,, \\
p^{\mu}&=p_0^\mu\,(1+\gamma p_0^{\rho}p_{0\rho})\,.
\end{align} 
\end{subequations}
From now on, the results will be presented in natural units, {\it i.e.} $c=\hbar=1$. With the parameters redefined as such, the position-momentum commutator is
\begin{equation}
    [x^{\mu},p^{\nu}]=i\,\left(1+\gamma p^{\rho}p_{\rho}\right)\eta^{\mu\nu}+2i\,\gamma p^{\mu}p^{\nu}\,.
\end{equation}
The Poincar\'e group corresponding to the RGUP above is represented by the following algebra
\begin{subequations}\label{Eq:QFTPoincare}
\begin{align}
   [x^\mu,M^{\nu\rho}] &=  i[1 + \gamma p^{\rho} p_{\,\rho}]\left(x^{\nu}\delta^{\mu\rho}-x^{\rho}\delta^{\mu\nu}\right) + i 2 \gamma p^{\mu} M^{\nu\rho}\,,\\
   [p^\mu, M^{\nu\rho}]& =  i[1 + \gamma  p^{\rho} p_{\,\rho}]\left(p^{\nu}\delta^{\mu\rho}-p^{\rho}\delta^{\mu\nu}\right)\,,\\
 [M^{\mu\nu},M^{\rho\sigma}]& = i\left(1 + \gamma p^{\rho} p_{\,\rho}\right)\left(\eta^{\mu\rho}M^{\nu\sigma}-\eta^{\mu\sigma} M^{\nu\rho} 
  -\eta^{\nu\rho}M^{\mu\sigma}+\eta^{\nu\sigma}M^{\mu\rho}\right)\,.
\end{align}
\end{subequations}
It is important to note that the deformed Poincar\'e algebra presented above has the same irreducible representations, and therefore describes the same particles as the standard Poincar\'e algebra. This feature can be seen derived in Appendix \ref{App:B}.
As we already showed, the physical momentum squared is a Casimir invariant of the modified Poincar\'e algebra presented above in Eq.\eqref{Eq:QFTPoincare}. Therefore, the Einstein dispersion relation for our case is 
\begin{equation}\label{Eq:QFTDispersionRelation}
   p^\mu p_\mu=-m^2\,,
\end{equation}
where $p_\mu$ is the physical momentum, the differential form of which 
is an equation of motion for any Lagrangian describing a boson field. The Dirac equations \textit{i.e.}, the equations of motion describing the dynamics of fermion fields, can also be easily derived.
Starting from the dispersion relation Eq.\eqref{Eq:QFTDispersionRelation} and  expressing it in terms of the auxiliary variables defined in Eq.\eqref{Eq:XPParticualrCase}, we get the modified Klein-Gordon equation in its differential form 
 \begin{equation}\label{Eq:DIfferentialKGEoM}
       \left[\partial_\mu\partial^\mu \left(1+ \gamma \partial_\nu \partial^\nu\right)^2
     + m\right] \phi = 0 \,.
 \end{equation}
Similarly, the differential form of the Dirac equation is shown to be 
 \begin{equation}\label{Eq:DiracEoM}
    \left(\tau^\mu p_\mu-m\right)\psi =\left[\tau^{\mu}p_{0\,\mu}(1+\gamma p_{0\rho}p_0^\rho)-m\right]\psi=0\,,
 \end{equation}
 where $\tau^\mu$ are the Dirac matrices and $\psi$ is a Dirac spinor. We can show that as part of the Lorentz group the generators of the Clifford algebra $Cl_{(3,1)}(\mathbb{R})$, if the Casimir is preserved  then the properties of the Clifford algebra will remain unchanged, as well. This is true in our case, thus, 
\begin{equation}\label{Eq:DifferentialDiracEoM}
     \left[ i\tau^\mu\partial_\mu(1+\gamma\partial_\rho\partial^\rho)^2 -m\right]\psi=0\,,
\end{equation}
and its Dirac conjugate
are the equations of motion for the RGUP-QED Lagrangian.
\section{RGUP modified Lagrangians}\label{sec:RGUPLagrangians}
  In the previous section, we established the modified equations of motion for the scalar, vector, and spinor fields. Note that the equations of motion are differential equations of higher than second order. Therefore, the Lagrangians corresponding to these equations of motion need to have higher than second derivative. The methodology of working with higher derivative Lagrangians
\begin{subequations}
\begin{align}
    \label{Eq:LagrangianGeneral}  L=&L(\phi,\dot\phi,\ddot\phi,\ldots,\overset{(n)}{\phi})\,,\\
  \label{Eq:LagrangianDensityGeneral}  \mathcal{L}=&\mathcal{L}(\phi,\partial_{\mu_1}\phi,\partial_{\mu_1}\partial_{\mu_2}\phi,\ldots,\partial_{\mu_1}\ldots\partial_{\mu_n}\phi)\,,
\end{align}
\end{subequations}
is given by the Ostrogradsky method \cite{Pons:1988tj,Woodard:2015zca,deUrries:1998obu}.The Euler-Lagrange equations for theories with higher derivatives will have the form:
\begin{equation}\label{Eq:OstrogradskyEulerLagrange}
    \frac{dL}{dq} -\frac{d}{dt}\frac{dL}{d\dot{q}}+\frac{d^2}{dt^2}\frac{dL}{d\ddot{q}}+\ldots+(-1)^n\frac{d^n}{dt^n}\frac{dL}{d (d^nq/dt^n)}=0\,,
\end{equation}
which in the case of fields is
\begin{equation}
    \frac{\partial\mathcal{L}}{\partial\phi}-  \partial_\mu  \frac{\partial\mathcal{L}}{\partial(\partial_\mu\phi)}+   \partial_{\mu_1} \partial_{\mu_2}\frac{\partial\mathcal{L}}{\partial(\partial_{\mu_1} \partial_{\mu_2}\phi)}+\ldots
    +(-1)^m\partial_{\mu_1}\ldots\partial_{\mu_m}\frac{\partial\mathcal{L}}{\partial(\partial_{\mu_1} \ldots\partial_{\mu_m}\phi)}=0\,.
\end{equation}
The Ostrogradsky method allows us to reconstruct the Lagrangian describing the dynamics of scalar and spinor fields from their equations of motion. 

\subsection{\label{sec:Lag}Scalar field Lagrangian}

We begin by assuming the most general form of higher derivative Lagrangian, which is one order higher than the equations of motion
\begin{align}\label{Eq:Lagrangian}
   \nonumber \mathcal{L}=\frac{1}{2}\partial_{\mu}\phi\partial^{\mu}\phi&+\gamma \left( C_1\,\partial_\mu\partial^\mu\phi\,\partial_\nu\partial^\nu\phi+C_2\,\partial_\mu\phi\,\partial^\mu\partial_\nu\partial^\nu\phi+C_3\,\partial_\nu\partial^\nu\partial^\mu\phi\,\partial_\mu\phi\right)\\\nonumber&+\gamma^2\left(C_4\, \partial_\mu\partial^\mu\partial_\nu\phi\,\partial^\nu\partial_\rho\partial^\rho\phi+C_5\, \partial_\mu\partial^\mu\partial_\nu\partial^\nu\phi\,\partial_\rho\partial^\rho\phi+C_6\, \partial_\mu\partial^\mu\partial_\nu\partial^\nu\partial_\rho\phi\,\partial^\rho\phi\right.\\
   &\left.+C_7\, \partial_\mu\partial^\mu\phi\,\partial_\nu\partial^\nu\partial_\rho\partial^\rho\phi+C_8\, \partial_\mu\phi\,\partial^\mu\partial_\nu\partial^\nu\partial_\rho\partial^\rho\phi\right)+C_9m^2\phi^2\,,
\end{align}
where $\partial_\mu = \partial / \partial x^\mu$. Using the Euler-Lagrange  equation prescribed by the Ostrogradsky method Eq.\eqref{Eq:OstrogradskyEulerLagrange}, we obtain the following Lagrangian 
\begin{equation}\label{Eq:RealLagrangian}
    \mathcal{L}_{\phi,\mathbb{R}}=\frac{1}{2}\partial_{\mu}\phi\partial^{\mu}\phi-\frac{1}{2}m^2\phi^2+\gamma \,\partial_\nu\partial^\nu\partial^\mu\phi\,\partial_\mu\phi
    +\frac{\gamma^2}{2}\partial_\mu\phi\,\partial^\mu\partial_\nu\partial^\nu\partial_\rho\partial^\rho\phi
    \,,
\end{equation}
As for the Lagrangian for a complex scalar field $\phi$, we generalize Eq.\eqref{Eq:RealLagrangian} by including additional terms obtaining 
\begin{multline}\label{Eq:ComplexLagrangian}
 \mathcal{L}_{\phi,\mathbb{C}} = \frac{1}{2} \left(\partial_{\mu}\phi\right)^\dagger \partial^{\mu}\phi - \frac{1}{2} m^2 \phi^\dagger \phi + \gamma \left[\left(\partial_\nu \partial^\nu \partial^\mu \phi \right)^\dagger \partial_\mu \phi + \partial_\nu \partial^\nu \partial^\mu \phi \left(\partial_\mu \phi \right)^\dagger \right]\\
+ \frac{\gamma^2}{2} \left[\left(\partial_\mu \phi \right)^\dagger \partial^\mu \partial_\nu \partial^\nu \partial_\rho \partial^\rho \phi + \partial_\mu \phi \left(\partial^\mu \partial_\nu \partial^\nu \partial_\rho \partial^\rho \phi \right)^\dagger \right],
\end{multline}
such that hermiticity is restored, \textit{i.e.} $\mathcal{L}_{\phi,\mathbb{C}}^{\dagger}=\mathcal{L}_{\phi,\mathbb{C}}$.
Furthermore, it is worth noticing that Eq.\eqref{Eq:ComplexLagrangian} is consistent with Eq.(55) in \cite{Kober:2010sj} up to a numerical factor.
\subsection{\label{sec:QED}Spinor field Lagrangian}
  The first step in the derivation of the spinor field Lagrangian is to assume a general form of the Lagrangian, where the order of derivatives is determined by the order of the equation of motion Eq.\eqref{Eq:DifferentialDiracEoM}.
Moreover, an  assumption is made that different terms will be multiplied by an arbitrary numerical coefficient 
\begin{equation}\label{Eq:GenearalDiracLagrangian}
     \mathcal{L}_{\psi}=\bar\psi\left[ iC_1\tau^\mu\partial_\mu(1+C_2\gamma\partial_\rho\partial^\rho) -C_3 m\right]\psi\,.
 \end{equation}
Next step is to prove that it has Eq.\eqref{Eq:DifferentialDiracEoM} as an equation of motion.
Applying the Ostrogradsky method, one gets the following equations of motion for the field and its complex conjugate
\begin{align}
     &C_1i\tau^\mu\partial_\mu\psi +C_1C_2\gamma\partial_\rho\partial^\rho\psi -C_3 m\psi=0,\\
     &C_1i\tau^\mu\partial_\mu\bar\psi+C_1C_2\gamma\partial_\rho\partial^\rho\bar\psi -C_3 m\bar\psi=0\,.
\end{align}
The equations of motion obtained through the Ostrogradsky method from Eq.\eqref{Eq:GenearalDiracLagrangian} need to be identical to Eq.\eqref{Eq:DifferentialDiracEoM}, which is  obtained from the dispersion relation.  
Therefore, the Lagrangian corresponding to the QFT spinor with minimum length will be of the form 
 \begin{equation}\label{Eq:DirackLagrangian}
     \mathcal{L}_{\psi}=\bar\psi\left[ i\tau^\mu\partial_\mu(1-\gamma\,\partial_\rho\partial^\rho) -m\right]\psi\,.
 \end{equation}

\subsection{$U(1)$ gauge field theory}
  The gauge field Lagrangian is obtained in a different way. The Ostrogradsky method is not utilized in this case. The derivation is  done by firstly assuming that the vector boson $A^\mu$ obeys the  ususal $U(1)$ gauge symmetries.
For the equations of motion, we  assume that they have the same RGUP corrections as the KG equation of the form Eq.\eqref{Eq:DIfferentialKGEoM}
 \begin{equation}\label{Eq:GaugeFieldEOM}
     \partial_\mu F^{\mu\nu}= \partial_\mu\partial^\mu A^\nu +2\gamma \partial_\mu\partial^\mu\partial_\rho\partial^\rho A^\nu  
     =0\,.
 \end{equation}
 Defining the standard gauge invariant
 field strength tensor as
 \begin{equation}
     F^{\mu\nu}_0=\partial^\mu A^\nu - \partial^\nu A^\mu,
 \end{equation}
one can express the RGUP modified field strength tensor in terms of the standard one up to first order in $\gamma$, as follows:
 \begin{equation}\label{Eq:EMFieldTensor}
    F^{\mu\nu}=F^{\mu\nu}_0+2\gamma\, \partial_\rho \partial^\rho F^{\mu\nu}_0\,.
 \end{equation}
 Then Eq.\eqref{Eq:GaugeFieldEOM} can be rewritten as 
 \begin{equation}\label{Eq:FieldTensorEoM}
    \partial_\mu F^{\mu\nu}=\partial_\mu F^{\mu\nu}_0+2\gamma  \partial_\rho \partial^\rho\partial_\mu F^{\mu\nu}_0
    +\gamma^2\partial_\sigma \partial^\sigma\partial_\rho \partial^\rho\partial_\mu F^{\mu\nu}_0
    \,.
 \end{equation}
 A test of gauge invariance of $F^{\mu\nu}$ is carried through by considering the following gauge transformation of the four-potential
 \begin{equation}
     A^\mu\rightarrow  A^{\prime\mu}=A^{\mu}+\partial^\mu\Lambda\,,
 \end{equation}
 %
 %
 %
and the field strength tensor is shown to be gauge invariant.
Thus, up to first order in the RGUP parameter $\gamma$, the gauge field Lagrangian reads
 \begin{equation}\label{Eq:GaugeFieldL}
     \mathcal{L}_{A}=-\frac{1}{4}F^{\mu\nu}F_{\mu\nu}=-\frac{1}{4}F^{\mu\nu}_0F_{\mu\nu 0}-\frac{\gamma }{2}F_{\mu\nu 0} \partial_\rho \partial^\rho F^{\mu\nu}_0
     \,.
 \end{equation}
Notice that both the field tensor in Eq.\eqref{Eq:EMFieldTensor} and the gauge field Lagrangian Eq.\eqref{Eq:GaugeFieldL} are invariant under $U(1)$ gauge transformations.

\section{Feynman rules}\label{Sec:FeynmanRules}
  We use the Lagrangians derived above to formulate a RGUP deformed scalar and spinor electrodynamics. This is achieved by  minimally coupling the lepton to the gauge fields.  The Feynman rules, consisting of the propagators and vertices for both cases, are  calculated using the standard methodology, which can be found in any QFT textbook.
  \subsection{Propagators}
  The Feynman propagator for the scalar field with a minimum length is the Green's functin of the   the modified KG equation in Eq.\eqref{Eq:DIfferentialKGEoM} 
\begin{equation}\label{GreenFunc}
     \left[\partial_\mu\partial^\mu \left(1+ \gamma \partial_\nu \partial^\nu\right)
     + (mc)^2 \right] G(x-x') = - i \delta(x-x')\,.
\end{equation}
Expressing the Green's function $G(x - x')$ in terms of its Fourier transform
\begin{equation}\label{GreenFourier}
    G(x-x')=\int \frac{d^4p_0}{(2\pi)^4} \tilde G(p_0) e^{-i p_0\cdot(x-x')},
\end{equation}
and substituting it in Eq.\eqref{GreenFunc}, we get 
\begin{equation}
\int \frac{d^4p_0}{(2\pi)^4} \tilde G(p_0) \left[-p_0^2 (1 - \gamma p_0^2) + (mc)^2\right] e^{-i p_0\cdot(x-x')}=-i \int\frac{d^4p_0}{(2\pi)^4}e^{-i p_0\cdot(x-x')}\,.
\end{equation}
The Fourier transform of the Feynman propagator is
\begin{equation}
     \tilde G(p_0)  =\frac{-i}{-p_0^2 (1+ \gamma p_0^2) + (mc)^2}\,,
\end{equation}
while the propagator itself is
\begin{equation}\label{Eq:ScalarPropagator}
     G(x-x')=\int \frac{d^4p_0}{(2\pi)^4} \frac{-i}{-p_0^2 ( 1 + \gamma p_0^2) + (mc)^2}e^{-i p_0\cdot(x-x')}\,.
\end{equation}
With the exception of a little trick consisting of 
multiplying both sides by 
\begin{equation}
    \left[\tau^{\mu}p_{0\,\mu}(1-\gamma p_{0\rho}p_0^\rho)+m\right]\,,
\end{equation}
we find the Dirac field propagator in the same way
\begin{equation}\label{Eq:DiracPropagator}
     G(x-x')=\int \frac{d^4p_0}{(2\pi)^4}\frac{-i \left[\tau^{\mu}p_{0\,\mu}(1+\gamma p_{0\rho}p_0^\rho)+m\right]}{p_0^{\mu}p_{0\,\mu}(1+\gamma p_{0\rho}p_0^\rho)^2-m^2}e^{-i p_0\cdot(x-x')}\,.
\end{equation}
Again, the gauge field propagator can be treated in a similar manner.
And  the Feynman propagator is found to be
\begin{equation}\label{Eq:GaugePropagator}
G(x-x')=\int \frac{d^4q_0}{(2\pi)^4} \frac{-i}{-q_0^2+2\gamma  q_0^4}e^{-i q_0\cdot(x-x')}\,,
\end{equation}
where $q_0$ is the auxiliary four-momentum of the gauge field.

This construction assumes that one can define Dirac delta functions in position space, and the minimal length arises when one tries to localize the fields. 

\subsection{Vertices}
  
The complete set of Feynman rules for the system requires the calculation of the vertices for the charged and gauge fields. Starting from the Lagrangian in Eq.\eqref{Eq:ComplexLagrangian}, one introduces 
the minimal coupling as follows 
\begin{equation}
     \partial_\mu \rightarrow D_\mu=\partial_\mu -ie A_\mu\,,
\end{equation}
where $A_\mu$ is the four-potential or the gauge field.

Replacing partials in Eq.\eqref{Eq:ComplexLagrangian} by covariant derivatives,   the full action of the minimally coupled complex scalar field and the gauge field is obtained. The action reads as follows
\begin{multline}
    \int \mathcal{L}\,d^4x = \int\left[\mathcal{L}_{A} + \mathcal{L}_{\phi,\mathbb{C}}\right]\,d^4x = \int\left\{\frac{1}{2} \left(D_{\mu} \phi\right)^\dagger D^{\mu} \phi - \frac{1}{2} m^2 \phi^\dagger \phi - \frac{1}{4}F^{\mu\nu}F_{\mu\nu}\right.\\
    + \gamma \left. \left[\left(D_\nu D^\nu D^\mu \phi\right)^\dagger D_\mu \phi + D_\nu D^\nu D^\mu \phi \left(D_\mu \phi\right)^\dagger \right] \right\}
+ \frac{\gamma^2}{2} \left[\left(D_\mu \phi\right)^\dagger D^\mu D_\nu D^\nu D_\rho D^\rho \phi\right.\\\left.\left. + D_\mu \phi \left(D^\mu D_\nu D^\nu D_\rho D^\rho \phi\right)^\dagger \right]\right\}
\,d^4x\, .
\end{multline}
By expanding  the Lagrangian we get a very rich theory
\begin{multline}\label{Eq:CoupledScalarLagrangian}
    \mathcal{L} = \frac{1}{2} \left(\partial_{\mu} \phi\right)^\dagger \partial^{\mu} \phi - i e A^\mu \left[\phi^\dagger \partial_\mu \phi - \phi \left(\partial_\mu \phi\right)^\dagger \right] - \frac{1}{2} m^2 \phi^\dagger \phi + e^2 A_\mu A^\mu \phi^\dagger \phi - \frac{1}{4} F^{\mu\nu}F_{\mu\nu}\\
    \gamma \left\{\left(\partial_\nu \partial^\nu \partial^\mu \phi\right)^\dagger \partial_\mu \phi + \partial_\nu \partial^\nu \partial^\mu \phi \left(\partial_\mu \phi\right)^\dagger - \frac{1}{4} F^{\mu\nu} F_{\mu\nu} \partial_\mu \partial_\nu F^{\mu\nu} \right. \\
    - i e \left\{\left(\partial_\nu \partial^\nu A^\mu\right) \left[\phi^\dagger \partial_\mu \phi - \phi \left(\partial_\mu\phi\right)^\dagger \right] + 2 \partial_\nu A^\mu \left[\left(\partial^\nu \phi\right)^\dagger \partial_\mu \phi - \partial^\nu \phi \left(\partial_\mu \phi^\dagger\right)\right] \right.\\
    \left. + A^\mu \left[\left(\partial_\nu \partial^\nu \phi\right)^\dagger \partial_\mu \phi - \partial_\nu \partial^\nu \phi \left(\partial_\mu \phi^\dagger\right)\right]\right\}
   + e^2 \left\{ A^\nu(\partial_\nu A_\mu) \left[\left(\partial^\mu \phi\right)^\dagger \phi + (\partial^\mu \phi) \phi^\dagger\right]\right.\\ + 2A^\nu A_\mu \left[\left(\partial^\mu \phi\right)^\dagger \partial_\nu \phi + \left(\partial^\mu \phi\right) \left(\partial_\nu \phi\right)^\dagger\right] \\
    + A^\nu A_\nu \left[ \left(\partial^\mu \phi\right)^\dagger \partial_\mu \phi + \left(\partial^\mu \phi\right) \left(\partial_\mu \phi\right)^\dagger\right] - 2 A^\mu A^\nu \left[\phi^\dagger\partial_\nu\partial_\mu\phi + \phi \left(\partial_\nu\partial_\mu\phi\right)^\dagger\right]\\
    \left. + 2 A^\mu (\partial_\nu \partial^\nu A_\mu) \phi^\dagger \phi + A^\mu (\partial^\nu A_\mu) \left[\phi^\dagger \partial_\nu \phi  + \phi \left(\partial_\nu \phi\right)^\dagger\right] + A^\mu A_\mu \left[\phi^\dagger \partial_\nu \partial^\nu \phi + \phi \left(\partial_\nu \partial^\nu \phi\right)^\dagger\right]\right\}\\
    + i e^3 \left\{ A_\mu A_\nu A^\nu \left[\phi \left(\partial^\mu \phi\right)^\dagger - \phi^\dagger\partial^\mu\phi\right] + 2 A^\mu A_\mu A^\nu \left[\phi^\dagger \partial_\nu \phi - \phi \left(\partial_\nu \phi\right)^\dagger\right] \right\} \\
    \left. + 2 e^4 A^\mu A_\mu A^\nu A_\nu\phi^\dagger\phi\right\} 
    + \mathcal{O}(\gamma^2)\,.
\end{multline}
The above expression contains all terms relevant to Feynman diagrams predicted by the usual scalar Quantum Electrodynamics (QED) Lagrangian with the addition of RGUP corrections.  By counting the fields in each term, it is easy to see that up to  6-point vertices (2 scalar and 4 gauge) are allowed. 
\begin{table}[ht]
\tbl{\label{Tbl:CouplingConstantTable}Classification of the Feynman vertices arising from Eq.\eqref{Eq:CoupledScalarLagrangian} in terms of the powers of the coupling constants: $\alpha$ the fine structure constant; $\gamma$ the minimum length coefficient. }
    {\begin{tabular}{@{}ccccc@{}}
\toprule
 &\multicolumn{4}{c}{\text{Powers of }$\gamma$} \\
 \hline\hline
   &$\alpha^{1/2}$&$\alpha^{1/2}\gamma$ &$\alpha^{1/2}\gamma^2$&$\alpha^{1/2}\gamma^3$ \\
 &$\alpha$&$\alpha\gamma$&$\alpha\gamma^2$&$\alpha\gamma^3$\\
  \text{Powers of $\alpha$}  & N/A &$\alpha^{3/2}\gamma$&$\alpha^{3/2}\gamma^2$ &$\alpha^{3/2}\gamma^3$  \\
 & N/A & $\alpha^{2}\gamma$&$\alpha^{2}\gamma^2$&$\alpha^{2}\gamma^3$ \\
   & N/A &$\alpha^{5/2}\gamma$&$\alpha^{5/2}\gamma^2$& $\alpha^{5/2}\gamma^3$ \\
  & N/A &$\alpha^3\gamma$&$\alpha^3\gamma^2$&$\alpha^3\gamma^3$\\
  \botrule
  \end{tabular}}
\end{table}
This Lagrangian contains 74 Feynman diagrams, with coupling constants classified by powers of the fine structure constant $\alpha$ and the minimum length parameter $\gamma$, see Table \ref{Tbl:CouplingConstantTable} . It is also worth mentioning that in the low energy limit $\gamma \rightarrow 0$  we recover the usual Lagrangian for a complex scalar fields.

Similar procedure is done for the Dirac field Lagrangian the
full RGUP-QED action  reads
\begin{align}
   S= \int \mathcal{L}\,d^4x &= \int\left[\mathcal{L}_{A} + \mathcal{L}_{\psi}\right]\,d^4x \nonumber\\ &= \int\bar{\psi}\left[ i\tau^\mu D_{\mu}\psi -\gamma \bar{\psi} i\tau^\mu D_{\mu}D_{\nu}D^{\nu}\psi  - \frac{1}{4}F^{\mu\nu}F_{\mu\nu}\right]\,\,d^4x \,.
   \label{qedaction1}
\end{align}
We can read off the vertices after we expand the covariant derivatives in the minimally coupled  modified Dirac field Lagrangian Eq.\eqref{Eq:DirackLagrangian} 
\begin{multline}\label{Eq:RGUPQED}
    \mathcal{L}_{\psi}=i\bar{\psi} \tau^\mu \partial_{\mu}\psi +i\gamma\bar{\psi}\tau^\mu \partial^{\rho}\partial_{\rho}\partial_{\mu}\psi-m\bar{\psi}\psi\\
    -e \left[\bar{\psi} \tau^\mu A_{\mu}\psi - 2 \gamma \bar{\psi} \tau^\mu \left(\partial_{\mu}A^{\rho}\right) \partial_\rho\psi - 2 \gamma\bar{\psi} \tau^\mu A^{\rho}\partial_{\mu}\partial_\rho\psi - \gamma\bar{\psi} \tau^\mu A_\mu\partial^{\rho}\partial_{\rho}\psi\right]\\
    - i e^2 \gamma \left[2 \bar{\psi} \tau^\mu A_\mu A^{\rho}\partial_{\rho}\psi + 2 \bar{\psi} \tau^\mu \left(\partial_\mu A^{\rho}\right)A_{\rho}\psi - \bar{\psi} \tau^\mu A^{\rho}A_{\rho}\partial_\mu\psi\right]\\
    - e^3\gamma\bar{\psi}\tau_{\mu}A^{\mu}A^{\rho}A_{\rho}\psi\,.
\end{multline}
Up to five particle vertices allowed,
they include always two fermions and from one to three gauge bosons.
\begin{table}[ht]
\tbl{\label{Tbl:CouplingConstantTable2}Classification of the coupling constants in Eq.\eqref{Eq:RGUPQED} in terms of the powers: $\alpha$ the fine structure constant; $\gamma$ the minimum length coefficient. }
{\begin{tabular}{@{}ccc@{}}
\toprule
   \quad  & \multicolumn{2}{c}{\text{Powers of }$\gamma$}\\  
 \hline\hline
  &$\alpha^{1/2}$&$\alpha^{1/2}\gamma$ \\
  \text{Powers of }$\alpha$ &N/A&$\alpha\gamma$\\
 & N/A & $\alpha^{3/2}\gamma$\\
  \botrule
  \end{tabular}}
\end{table}
Furthermore,  the coupling constants for each vertex are a product of powers of the electronic charge $e$ and the RGUP coefficient $\gamma$.
In fact, the power of the electronic charge determines how many bosons couple to the vertex and the power of $\gamma$ is $0$ for the usual terms and $1$ for the RGUP corrections terms. As before we can classify the coupling constants in Table \ref{Tbl:CouplingConstantTable2}.

\section{RGUP Corrections to QED scattering amplitudes}\label{sec:Crosssections}
The calculations for the scattering amplitudes can be quite cumbersome so we will not present them in full here. If the reader is interested the full calculations are presented in \cite{Todorinov2020,Bosso2020-dv,Bosso2020-fz}. 

The particular case we will be studying is an electron-muon scattering. We made this choice to avoid mass degeneracies which will complicate our calculations without helping in clarifying the concept. First when calculating the scattering amplitudes the focus will be on the 3-point vertices, containing up to first order the RGUP coefficient $\gamma$. This approximation is justified by the fact that the 3-point vertices will have the largest contribution to the scattering amplitudes.

In brief, we begin by calculating the transition amplitudes for a single three legged vertex. 
For the scalar field  we get 
\begin{subequations} \label{Eq:AmplitudeSeparated}
\begin{align}
 \nonumber T_{fi}=&-i \int e A^\mu\left[\phi_f^\dagger \partial_\mu\phi_i-\phi_f\partial_\mu\phi_i^\dagger \right] d^4x\\
 \nonumber&-i \int e\gamma  A^\mu\left[4\partial_\nu\partial^\nu\phi^\dagger \partial_\mu\phi+4\partial^\nu\phi^\dagger \partial_\nu\partial_\mu\phi+\phi^\dagger \partial_\nu\partial^\nu\partial_\mu\phi-4\partial_\nu\partial^\nu\phi\partial_\mu\phi^\dagger \right. \\
 &\left.-4\partial^\nu\phi\partial_\nu\partial_\mu\phi^\dagger -\phi\partial_\nu\partial^\nu\partial_\mu\phi^\dagger  \right]d^4x\,.
\end{align}
\end{subequations}
In the QED case we get for the different components
 \begin{subequations} \label{Eq:DiracAmplitudeSeparated}
\begin{align}
 \nonumber T_{fi}&=i\int e\bar{\psi}_f \tau^\mu A_{\mu}\psi_i\, d^4 x+i\int2e\gamma\partial_{\rho}\bar{\psi}_f \tau^\rho A^{\mu}\partial_\mu\psi_i\, d^4 x\,,\\
 &-i\int e\gamma\bar{\psi} \tau^\mu A_\mu\partial^{\rho}\partial_{\rho}\psi\, d^4 x\,.
\end{align}
\end{subequations}
The separation in different terms is purely for convenience.

We then couple two Feynman vertices thorough a gauge boson $A^\mu$, and isolate the modified invariant amplitude squared
\begin{subequations}
\begin{align}
   \left\vert\mathfrak{M}_{\text{Scalar}}\right\vert^2&=4\pi  \alpha \frac{3+\cos\theta}{1-\cos\theta}\left[1+8\gamma E^2(1-\cos\theta)\right]\,,\\
       \left|\mathfrak{M}_{\text{spinor}}\right|^2&=32\pi^2\epsilon_0^2\alpha\left[\frac{t^2+u^2}{s^2}+ \frac{1}{2}\gamma(m_e^2+m_{\text{muon}}^2)\frac{tu-u^2}{s^2} \right]\,,
\end{align}
\end{subequations}
where $s,t,u$ are the Mandelstam variables given by 
\begin{subequations}
\begin{align}
    s&=\left(k+k'\right)^2=\left(p+p'\right)^2\approx2k\cdot k'\approx 2p\cdot p'\,,\\
    t&=\left(k-p\right)^2=\left(p'-k'\right)^2\approx -2k\cdot p\approx -2k'\cdot p'\,,\\
    u&=\left(k-p'\right)^2=\left(p-k'\right)\approx -2k\cdot p'\approx -2p\cdot k'\,.
\end{align}
\end{subequations}
From here, it is easy to calculate the RGUP scaled transition rates, from which we can isolate a formula for the differential cross-ectons for both cases. For the scalar fields, the differential cross-section is 
\begin{equation}\label{Eq:FinalScalarDIferentialCrossSection}
    \left.\frac{d\sigma}{d\Omega}\right\vert_{CM} = \frac{1}{4\,s}\alpha^2\left(\frac{3+\cos\theta}{1-\cos\theta}\right)^2\left[1+16\gamma E^2(1-\cos\theta)\right]\,,
\end{equation}
and for the spinors
\begin{equation}\label{Eq:FinalDIferentialCrossSection}
\left. \frac{d\sigma}{d\Omega} \right\vert_{CM} = \frac{\alpha^2}{4\,s} \left[ \frac{1}{2} \left(1 + \cos^2\theta\right) + \frac{1}{4} \gamma (m_e^2 + m_{\text{muon}}^2) \left(\cos\theta+\cos^2\theta\right)\right]\,.
 \end{equation}
 As one can see from the differential cross-section for the scalar fields Eq. \eqref{Eq:FinalScalarDIferentialCrossSection}, the magnitude og QG effects depends on the energy of the scattering squared, and it will be largest for the backscattering.  
 We can see that this is also the case in the spinor case  Eq.\eqref{Eq:FinalDIferentialCrossSection}. It is interesting to note that for the spinor case the corrections depend on the mass squared of the particles involved.
This suggests that minimum length effects on scattering amplitudes may be measurable for electromagnetic scattering of heavier systems, such as scattering of heavy ions. Applied to existing data from the Xe-Xe scattering observed in the ATLAS experiment, the order of magnitude of the corrections is $\gamma\, m_{\text{Xe}}^2 /4\sim 10^{-34}$ \cite{Balek:2019nqk}. The respective upper bounds on the RGUP parameter are then set to be $\gamma_0<10^{34}$.
\section{Conclusion}
In this work we have extended the Generalized Uncertainty Principle to relativistic framework. This leads to the solution of two major problems: the dependence of the minimum measurable length on the frame of reference; and the composition law problem.  This allowed us to extend the reach of GUP from experiments and effects in Quantum Mechanics into Quantum Fired Theory. We achieved that through the formulation of Effective Field Theory with minimum length. Further, we were able to calculate the scattering amplitudes for high-enegy scattering of scalar and spinor particles. Comparing our results to data collecd in the ATLAS experiments, we were able to obtain bounds on the RGUP parameters, \textit{i.e.} $\gamma_0<10^{34}$. This bound does not seem impressive on the first glance, however, it is sixteen orders of magnitude improvement compared to previous results from purely quadratic GUP models. 

Additionally, we have already been able to obtain RGUP modified action for a spin-2 particle, and through the classical limit connect it to quadratic gravity, allowing us to test our ideas into fields like cosmological inflation. 

The results presented here provide a valuable step to extend the reach of Quantum Gravity Phenomenology to the most highly energetic earthbound experiments like LHC. For example, we need to  generalize the results here for the gauge group of the Standard Model $U(1)\otimes SU(2)\otimes SU(3)$. Additionally, the gauge field Lagrangian can be used to calculate corrections to phenomena in classical electrodynamics and light propagation, amongst others, like cosmology with weakly self-interacting photons, gravitational waves, and Higgs mechanism and mass. 

\appendix{Ostrogradsky method and Vacuum Instability}\label{App:A}
The Ostrogradsky method is a mathematical constructions allowing one to work with higher derivative Lagrangians \cite{Pons:1988tj,Woodard:2015zca,deUrries:1998obu}
. One of the major issues with it, is the instability of the Hamiltonian, arising when one considers higher than second order derivative. The instability for the particular case presented in here is discussed in the following appendix. We show that the Hamiltonian for a scalar field is positively defined in sub Planckian regimes, thus fixing the applicable domain for our Effective field theory.

 We  from the equation of motion for the real scalar field, obtained from the modified dispersion relation which corresponds to the Casimir operator of the modified Poincar\'e group defined in \cite{Todorinov2018-xi} 
 \begin{equation}\label{ModKG}
     p_0^{\rho}p_{0\rho}(1+2\alpha\gamma^2p_0^{\sigma}p_{0\sigma})=-(mc)^2 \,.
 \end{equation}
  The Lagrangian for a real scalar field is derived by assuming the most general form of the Lagrangian, and applying the Ostrogradsky method to obtain its equations of motion and comparing to Eq.\eqref{ModKG}.
  The form of the Lagrangian is
\begin{equation}\label{RealL}
    \mathcal{L}_{\phi,\mathbb{R}}=\frac{1}{2}\partial_{\mu}\phi\partial^{\mu}\phi-\frac{1}{2}m^2\phi^2+\gamma \,\partial_\nu\partial^\nu\partial^\mu\phi\,\partial_\mu\phi
    \,.
\end{equation}
The generalized coordinates  are defined as 
\begin{equation}\label{generalizedfieldcoordinates1}
    q_1\equiv\phi\quad     q_2\equiv\partial_\mu\partial^\mu\phi\,.
\end{equation}
The generalized momenta are then derived as
\begin{subequations}
\begin{align}
     \pi^{{\mu}\rho}_\rho&=
   \frac{\partial\mathcal{L}}{\partial(\partial_\mu\partial_\rho \partial^\rho \phi)}=2\gamma\partial^\mu \phi\,,\\
   \pi^{\mu}&=
   \frac{\partial\mathcal{L}}{\partial(\partial_\mu \phi)}-\partial_\sigma\partial^\sigma\pi^{\mu\rho}_\rho=\partial^\mu \phi\,.
\end{align}
\end{subequations}
The Hamiltonian density is then obtained using the Ostrogradsky Legendre transformations
\begin{subequations}
\begin{align}
    \mathcal{H}&=\pi^\mu\partial_\mu\phi+\pi^{\mu\rho}_\rho\partial_\sigma\partial^\sigma\partial_\mu\phi-\frac{1}{2}\pi^\mu\pi_\mu -\pi^{\mu\rho}_\rho\partial^\sigma\partial_\sigma\pi_\mu+\frac{1}{2}m\phi^2\\
    &= \frac{1}{2}\pi^\mu\pi_\mu+\frac{1}{2}m^2\phi^2\,.
\end{align}
\end{subequations}
We can easily see that the above is positive definite. However, we can also see that the correction terms cancel, the Quantum Gravity corrections are hidden in the dispersion relation of $\phi$. The field $\phi$ is solution of Eq.\eqref{ModKG} which has four different solutions all containing  Quantum Gravity corrections. The calculation above is done in the framework of De Donder Weyl Covariant Hamiltonian
Formulation of Field Theory.

Considering the same definition of generalized coordinates presented in Eq.\eqref{generalizedfieldcoordinates1}, one can derive the field momenta outside of the De Donder Weyl formulation as 
\begin{subequations}
\begin{align}
     \pi^{\rho}_\rho&=
   \frac{\partial\mathcal{L}}{\partial(\partial_\rho \partial^\rho \dot\phi)}=2\gamma\dot\phi\,,\\
   \pi&=
   \frac{\partial\mathcal{L}}{\partial( \dot\phi)}-\partial_\sigma\partial^\sigma\pi^{\rho}_\rho=\dot\phi\,.
\end{align}
\end{subequations}
 Performing the Legendre transformation the Hamiltonian density is obtained as
 \begin{align}\label{realscalarfield}
    \mathcal{H}&=\pi\dot\phi+\pi^{\rho}_\rho\partial_\sigma\partial^\sigma\dot\phi-\frac{1}{2}\pi\pi -\pi^{\rho}_\rho\partial^\sigma\partial_\sigma\pi+\frac{1}{2}m\phi^2\nonumber\\&+\frac{1}{2}\left(\nabla \phi\right)\cdot\left(\nabla \phi\right)+\gamma\left(\nabla \phi\right)\cdot\left(\partial_\sigma\partial^\sigma\nabla \phi\right)\\
    &= \frac{1}{2}\pi\pi+\frac{1}{2}m\phi^2+\frac{1}{2}\left(\nabla \phi\right)\cdot\left(\nabla \phi\right)+\gamma\left(\nabla \phi\right)\cdot\left(\partial_\sigma\partial^\sigma\nabla \phi\right)\,.
\end{align}
We can not be sure if he Hamiltonian density is positive definite. This is due to the fact that one is not sure of the sign of the term proportional to $\gamma=\frac{\gamma_0}{(M_{\mathrm{Pl}}\,c)^2}$. If one assumes that $\gamma_0$ is of the order one, the coefficient is $\gamma\sim 10^{-38}$. Therefore, for energies smaller than the Planck energy $E<E_{Pl}$, the leading order terms in  Eq.\eqref{realscalarfield} are several orders of magnitude bigger than the quantum gravity corrections, which means that the Hamiltonian density on the whole is positive definite and does \textit{not} have Ostrogradsky instability. Conclusion about the positive definition of the Hamiltonian for energies bigger than the Planck energy $E>E_{Pl}$ cannot be drawn. Making the QFT presented here an effective theory.

\appendix{Modified Poncar\'e algebra representations}\label{App:B}
Let us explore the RGUP modified Poincar\'e algebra.
The physical position $x_\mu$ and momentum $p_\mu$ are functions of the auxiliary ones represented as 
 \begin{subequations}
\begin{align}
x^{\mu}&=x_0^{\mu} \\
p^{\mu}&=p_0^\mu\,(1+\gamma p_0^{\rho}p_{0\rho})\,.
\end{align} 
\end{subequations}
While the Lorentz generators are defined as 
\begin{equation}
    M^{\mu\nu} = p^{\mu}x^{\nu}-p^{\nu}x^{\mu}
    = \left[1+\gamma p_0^{\rho}p_{0\,\rho}\right]\tilde{M}^{\mu\nu}\,,
\end{equation}
where it is wort mentioning that all the results are truncated to first order in the RGUP parameter $\gamma$.
The modified Poincar\'e algebra is then calculated to be %
\begin{subequations}\label{Eq:PoincareAlgebra}
\begin{align}
\label{Eq:xM}  [x^\mu,M^{\nu\rho}] &=  i\hbar[1 +  \gamma p^{\rho} p_{\,\rho}]\left(x^{\nu}\delta^{\mu\rho}-x^{\rho}\delta^{\mu\nu}\right) + i\hbar 2\gamma p^{\mu} M^{\nu\rho}\\
   \label{Eq:pM}[p^\mu, M^{\nu\rho}]& =  i\hbar[1 +  \gamma p^{\rho} p_{\,\rho}]\left(p^{\nu}\delta^{\mu\rho}-p^{\rho}\delta^{\mu\nu}\right),\\
\label{Eq:MM}  [M^{\mu\nu},M^{\rho\sigma}]& = i\hbar\left(1 +\gamma p^{\rho} p_{\,\rho}\right)\left(\eta^{\mu\rho}M^{\nu\sigma}-\eta^{\mu\sigma} M^{\nu\rho} 
  -\eta^{\nu\rho}M^{\mu\sigma}+\eta^{\nu\sigma}M^{\mu\rho}\right)\,.
\end{align}
\end{subequations}
We can show that the physical rotations $J_i$ and boosts $K_i$ are as follows 
\begin{subequations}
\begin{align}
J_i&=\frac{1}{2}\varepsilon_{imn}M^{mn}=\frac{1}{2}\left(1 +\gamma p^{\rho} p_{\,\rho}\right)\varepsilon_{imn}\tilde{M}^{mn}\\
K_i&=M_{0i}=\left(1 +\gamma p^{\rho} p_{\,\rho}\right)\tilde{M}_{0i}\,,
\end{align}
\end{subequations}
The algebra for which is 
 \begin{subequations}
\begin{align}
[J_i,J_j]&=-i\varepsilon_{ijk}\left(1 +\gamma p^{\rho} p_{\,\rho}\right)J^k\,, \\
[K_i,K_j]&=i\varepsilon_{ijk}\left(1 +\gamma p^{\rho} p_{\,\rho}\right)J^k \,,\\
[J_i,K_j]&=i\varepsilon_{ijk}\left(1 +\gamma p^{\rho} p_{\,\rho}\right)K^k\,.
\end{align} 
\end{subequations}
Once again we can define new set of this time physical operators $A_i$ and $B_i$
 \begin{subequations}
\begin{align}
A_i=\frac{1}{2}\left(J_i+i K_i\right)=\frac{1}{2}\left(1 +\gamma p^{\rho} p_{\,\rho}\right)\left(\tilde{J}_i+i\tilde{K}_i\right)\\
B_i=\frac{1}{2}\left(J_i-i K_i\right)=\frac{1}{2}\left(1 +\gamma p^{\rho} p_{\,\rho}\right)\left(\tilde{J}_i-i\tilde{K}_i\right)\,.
\end{align} 
\end{subequations}
Which form the following algebra
 \begin{subequations}
\begin{align}
[A_i,A_j]&=-i\varepsilon_{ijk}\left(1 +\gamma p^{\rho} p_{\,\rho}\right)A^k\,, \\
[B_i,B_j]&=-i\varepsilon_{ijk}\left(1 +\gamma p^{\rho} p_{\,\rho}\right)B^k \,,\\
[A_i,B_j]&=0\,.
\end{align} 
\end{subequations}
We notice that once again this algebra is equivalent to $SU(2)\otimes SU(2)$ algebra. The irreducible representation of which are scalar $(0,0)$, spinors/fermions $(0,1/2)$ and  $(1/2,0)$, vector bosons/ photons $(1/2,1/2)$, and potentially gravitons $(1,1)$.
We have taken two things in consideration: First that the modified Poincar\'e algebra has both the auxiliary momentum squared $p_{0\,\mu}p_0^\mu$ and the physical momentum squared  $p_{\mu}p^\mu$ as Casimir invariants; And second that those quantities are scalars and therefore modify the structure constants of the algebra without actually changing its physical properties.

\bibliographystyle{ws-procs961x669}
\bibliography{ws-pro-sample}

\begin{thebibliography}{10}

\bibitem{Kempf:1994su}
A.~Kempf, G.~Mangano and R.~B. Mann, {Hilbert space representation of the
  minimal length uncertainty relation}, {\em Phys. Rev. D} {\bf 52}, 1108
  (1995).

\bibitem{Adler1999-db}
R.~J. Adler and D.~I. Santiago, On gravity and the uncertainty principle, {\em
  Mod. Phys. Lett. A} {\bf 14}, p. 1371 (April 1999).

\bibitem{Adler_2001}
R.~J. Adler, P.~Chen and D.~I. Santiago, The generalized uncertainty principle
  and black hole remnants, {\em Gen. Relat. Grav.} {\bf 33}, p. 2101–2108
  (Dec 2001).

\bibitem{Ali2011}
A.~F. Ali, S.~Das and E.~C. Vagenas, {A proposal for testing Quantum Gravity in
  the lab}, {\em Phys. Rev. D} {\bf 84}, p. 044013  (2011).

\bibitem{Ali_2014}
A.~F. Ali and B.~Majumder, Towards a cosmology with minimal length and maximal
  energy, {\em Class. Quant. Grav.} {\bf 31}, p. 215007 (Oct 2014).

\bibitem{Ali_2015}
A.~F. Ali, M.~Faizal and M.~M. Khalil, Short distance physics of the
  inflationary de sitter universe, {\em JCAP} {\bf 2015}, p. 025–025 (Sep
  2015).

\bibitem{Alonso_Serrano_2018}
A.~Alonso-Serrano, M.~P. Dabrowski and H.~Gohar, Generalized uncertainty
  principle impact onto the black holes information flux and the sparsity of
  hawking radiation, {\em Phys. Rev. D} {\bf 97} (Feb 2018).

\bibitem{Amati1989-gs}
D.~Amati, M.~Ciafaloni and G.~Veneziano, Can spacetime be probed below the
  string size?, {\em Phys. Lett. B} {\bf 216}, 41 (January 1989).

\bibitem{Amelino_Camelia_2001}
G.~Amelino-Camelia, Testable scenario for relativity with minimum length, {\em
  Phys. Lett. B} {\bf 510}, p. 255–263 (Jun 2001).

\bibitem{AMELINO_CAMELIA_2002}
G.~Amelino-Camelia, Relativity in spacetimes with short-distance structure
  governed by an observer-independent (planckian) length scale, {\em Int. J.
  Mod. Phys. D} {\bf 11}, p. 35–59 (Jan 2002).

\bibitem{Amelino-Camelia2013-xs}
G.~Amelino-Camelia, {Quantum-Spacetime} phenomenology, {\em Living Rev.
  Relativ.} {\bf 16}, p.~5 (June 2013).

\bibitem{Bargue_o_2015}
P.~Bargueño and E.~C. Vagenas, Semiclassical corrections to black hole entropy
  and the generalized uncertainty principle, {\em Phys. Lett. B} {\bf 742}, p.
  15–18 (Mar 2015).

\bibitem{Bambi2007-te}
C.~Bambi and F.~R. Urban, Natural extension of the generalised uncertainty
  principle, {\em Class. Quant. Grav.} {\bf 25}, p. 095006 (September 2007).

\bibitem{Bawaj_2015}
M.~Bawaj and et~al., Probing deformed commutators with macroscopic harmonic
  oscillators, {\em Nature Communications} {\bf 6} (Jun 2015).

\bibitem{Bojowald2011-bb}
M.~Bojowald and A.~Kempf, Generalized uncertainty principles and localization
  of a particle in discrete space, {\em Phys. Rev. D} {\bf 86}, p. 085017
  (December 2011).

\bibitem{Bolen2005-jq}
B.~Bolen and M.~Cavagli{\'a}, (anti-)de sitter black hole thermodynamics and
  the generalized uncertainty principle, {\em Gen. Relat. Grav.} {\bf 37}, 1255
  (July 2005).

\bibitem{bosso2017generalized}
P.~Bosso and S.~Das, Generalized uncertainty principle and angular momentum,
  {\em Ann. of Phys.} {\bf 383}, 416  (2017).

\bibitem{Bosso2018}
P.~Bosso, S.~Das and R.~B. Mann, {Potential tests of the Generalized
  Uncertainty Principle in the advanced LIGO experiment}, {\em Phys. Lett. B}
  {\bf 785}, 498  (2018).

\bibitem{Bosso:2018uus}
P.~Bosso, {Rigorous Hamiltonian and Lagrangian analysis of classical and
  quantum theories with minimal length}, {\em Phys. Rev. D} {\bf 97}, p. 126010
   (2018).

\bibitem{Bosso:2019ljf}
P.~Bosso and O.~Obreg\'on, {Minimal length effects on quantum cosmology and
  quantum black hole models}, {\em Class. Quant. Grav.} {\bf 37}, p. 045003
  (2020).

\bibitem{Burger_2018}
D.~J. Burger and et~al, Towards the raychaudhuri equation beyond general
  relativity, {\em Phys. Rev. D} {\bf 98} (Jul 2018).

\bibitem{bushev2019testing}
P.~Bushev and et~al, Testing the generalized uncertainty principle with
  macroscopic mechanical oscillators and pendulums, {\em Phys. Rev. D} {\bf
  100}, p. 066020  (2019).

\bibitem{Capozziello:1999wx}
S.~Capozziello, G.~Lambiase and G.~Scarpetta, {Generalized uncertainty
  principle from quantum geometry}, {\em Int. J. Theor. Phys.} {\bf 39}, 15
  (2000).

\bibitem{Casadio_2020}
R.~Casadio and F.~Scardigli, Generalized uncertainty principle, classical
  mechanics, and general relativity, {\em Phys. Lett. B} {\bf 807}, p. 135558
  (Aug 2020).

\bibitem{Chang:2011jj}
L.~N. Chang and et~al, {On the Minimal Length Uncertainty Relation and the
  Foundations of String Theory}, {\em Adv. High Energy Phys.} {\bf 2011}, p.
  493514  (2011).

\bibitem{Cortes:2004qn}
J.~L. Cortes and J.~Gamboa, {Quantum uncertainty in doubly special relativity},
  {\em Phys. Rev. D} {\bf 71}, p. 065015  (2005).

\bibitem{Costa_Filho2016-ox}
R.~N. Costa~Filho and et~al, Extended uncertainty from first principles, {\em
  Phys. Lett. B} {\bf 755}, 367 (April 2016).

\bibitem{Dabrowski2019-cb}
M.~P. Dabrowski and F.~Wagner, {Extended Uncertainty Principle for Rindler and
  cosmological horizons}, {\em Eur. Phys. J. C} {\bf 79}, p. 716  (2019).

\bibitem{Dabrowski2020-kk}
M.~P. Dabrowski and F.~Wagner, Asymptotic generalized extended uncertainty
  principle, {\em Eur. Phys. J. C} {\bf 80}, p. 676 (June 2020).

\bibitem{Das2008}
S.~Das and E.~C. Vagenas, {Universality of Quantum Gravity Corrections}, {\em
  Phys. Rev. Lett.} {\bf 101}, p. 221301  (2008).

\bibitem{Das:2009hs}
S.~Das and E.~C. Vagenas, {Phenomenological Implications of the Generalized
  Uncertainty Principle}, {\em Can. J. Phys.} {\bf 87}, 233  (2009).

\bibitem{Das:2010zf}
S.~Das, E.~C. Vagenas and A.~F. Ali, {Discreteness of Space from GUP II:
  Relativistic Wave Equations}, {\em Phys. Lett. B} {\bf 690}, 407  (2010),
  [Erratum: Phys.Lett.B 692, 342--342 (2010)].

\bibitem{DAS2011596}
S.~Das and R.~Mann, Planck scale effects on some low energy quantum phenomena,
  {\em Phys. Lett. B} {\bf 704}, 596  (2011).

\bibitem{Das2014}
S.~Das, M.~P.~G. Robbins and M.~A. Walton, {Generalized Uncertainty Principle
  Corrections to the Simple Harmonic Oscillator in Phase Space}, {\em Can. J.
  Phys.} {\bf 94}, 139  (2016).

\bibitem{Das_2019}
S.~Das, S.~S. Haque and B.~Underwood, Constraints and horizons for de sitter
  with extra dimensions, {\em Phys. Rev. D} {\bf 100} (Aug 2019).

\bibitem{Das_2020}
A.~Das, S.~Das and E.~C. Vagenas, Discreteness of space from gup in strong
  gravitational fields, {\em Phys. Lett. B} {\bf 809}, p. 135772 (Oct 2020).

\bibitem{Garcia-Chung:2020zyq}
A.~Garcia-Chung and et~al, {Propagation of quantum gravity-modified
  gravitational waves on a classical FLRW spacetime}, {\em Phys. Rev. D} {\bf
  103}, p. 084053  (2021).

\bibitem{Giddings2020-xz}
S.~B. Giddings, Black holes and other clues to the quantum structure of
  gravity, {\em Galaxies} {\bf 9}, p.~16 (December 2020).

\bibitem{Hamil2019-qh}
B.~Hamil, M.~Merad and T.~Birkandan, Applications of the extended uncertainty
  principle in {AdS} and ds spaces, {\em Eur. Phys. J. Plus} {\bf 134}, p. 278
  (June 2019).

\bibitem{Hossenfelder:2006cw}
S.~Hossenfelder, {Interpretation of quantum field theories with a minimal
  length scale}, {\em Phys. Rev. D} {\bf 73}, p. 105013  (2006).

\bibitem{Hossenfelder:2012jw}
S.~Hossenfelder, {Minimal Length Scale Scenarios for Quantum Gravity}, {\em
  Living Rev. Rel.} {\bf 16}, p.~2  (2013).

\bibitem{Kober:2010sj}
M.~Kober, {Gauge Theories under Incorporation of a Generalized Uncertainty
  Principle}, {\em Phys. Rev. D} {\bf 82}, p. 085017  (2010).

\bibitem{KONISHI1990276}
K.~Konishi, G.~Paffuti and P.~Provero, Minimum physical length and the
  generalized uncertainty principle in string theory, {\em Phys. Lett. B} {\bf
  234}, 276   (1990).

\bibitem{MAGGIORE199365}
M.~Maggiore, A generalized uncertainty principle in quantum gravity, {\em Phys.
  Lett. B} {\bf 304}, 65   (1993).

\bibitem{Maggiore:1994}
M.~Maggiore, Quantum groups, gravity, and the generalized uncertainty
  principle, {\em Phys. Rev. D} {\bf 49}, 5182 (May 1994).

\bibitem{Marin:2013pga}
F.~Marin {\em et~al.}, {Gravitational bar detectors set limits to Planck-scale
  physics on macroscopic variables}, {\em Nature Phys.} {\bf 9}, 71  (2013).

\bibitem{Mead1966-xj}
C.~A. Mead, Observable consequences of {Fundamental-Length} hypotheses, {\em
  Phys. Rev.} {\bf 143}, 990 (March 1966).

\bibitem{Moradpour2021-jy}
H.~Moradpour, S.~Aghababaei and A.~H. Ziaie, A note on effects of generalized
  and extended uncertainty principles on j{\"u}ttner gas, {\em Symmetry} {\bf
  13}, p. 213 (January 2021).

\bibitem{Mureika2019-lf}
J.~R. Mureika, Extended uncertainty principle black holes, {\em Phys. Lett. B}
  {\bf 789}, 88 (February 2019).

\bibitem{Myung_2007}
Y.~S. Myung, Y.-W. Kim and Y.-J. Park, Black hole thermodynamics with
  generalized uncertainty principle, {\em Phys. Lett. B} {\bf 645}, p.
  393–397 (Feb 2007).

\bibitem{Park2008-uj}
M.-I. Park, The generalized uncertainty principle in ({A)dS} space and the
  modification of hawking temperature from the minimal length, {\em Phys. Lett.
  B} {\bf 659}, 698 (January 2008).

\bibitem{Scardigli1999-ne}
F.~Scardigli, Generalized uncertainty principle in quantum gravity from micro -
  black hole gedanken experiment, {\em Phys. Lett. B} {\bf 452}, 39 (April
  1999).

\bibitem{Snyder:1946qz}
H.~S. Snyder, Quantized space-time, {\em Phys. Rev.} {\bf 71}, 38 (Jan 1947).

\bibitem{Sprenger_2011}
M.~Sprenger, P.~Nicolini and M.~Bleicher, Neutrino oscillations as a novel
  probe for a minimal length, {\em Class. Quant. Grav} {\bf 28}, p. 235019 (Nov
  2011).

\bibitem{Sriramkumar:2006qt}
L.~Sriramkumar and S.~Shankaranarayanan, {Path integral duality and Planck
  scale corrections to the primordial spectrum in exponential inflation}, {\em
  JHEP} {\bf 12}, p. 050  (2006).

\bibitem{Stargen_2019}
D.~J. Stargen, S.~Shankaranarayanan and S.~Das, Polymer quantization and
  advanced gravitational wave detector, {\em Phys. Rev. D} {\bf 100} (Oct
  2019).

\bibitem{tedesco2011fine}
L.~Tedesco, Fine structure constant, domain walls, and generalized uncertainty
  principle in the universe, {\em Int. J. Math. and Math. S.} {\bf 2011}
  (2011).

\bibitem{wang2016solutions}
B.~Wang, C.~Long, Z.~Long and T.~Xu, Solutions of the schr{\"o}dinger equation
  under topological defects space-times and generalized uncertainty principle,
  {\em Eur. Phys. J. Plus} {\bf 131}, p. 378  (2016).

\bibitem{Hossenfelder:2014ifa}
S.~Hossenfelder, {The Soccer-Ball Problem}, {\em SIGMA} {\bf 10}, p. 074
  (2014).

\bibitem{Amelino-Camelia:2014gga}
G.~Amelino-Camelia, {Planck-scale soccer-ball problem: a case of mistaken
  identity}, {\em Entropy} {\bf 19}, p. 400  (2017).

\bibitem{Lake:2019oaz}
M.~J. Lake, {A Solution to the Soccer Ball Problem for Generalized Uncertainty
  Relations}, {\em Ukr. J. Phys.} {\bf 64}, 1036  (2019).

\bibitem{Todorinov2018-xi}
V.~Todorinov, P.~Bosso and S.~Das, Relativistic generalized uncertainty
  principle, {\em Ann. Phys.} {\bf 405}, 92 (October 2018).

\bibitem{Quesne2006}
C.~Quesne and V.~Tkachuk, {Lorentz-covariant deformed algebra with minimal
  length}, {\em Czech. J. Phys.} {\bf 56}, 1269  (2006).

\bibitem{Bosso2020-dv}
P.~Bosso, S.~Das and V.~Todorinov, Quantum field theory with the generalized
  uncertainty principle i: Scalar electrodynamics, {\em Ann. Phys.} {\bf 422},
  p. 168319 (May 2020).

\bibitem{Bosso2020-fz}
P.~Bosso, S.~Das and V.~Todorinov, Quantum field theory with the generalized
  uncertainty principle {II}: Quantum electrodynamics, {\em Ann. Phys.} {\bf
  424}, p. 168350 (May 2020).

\bibitem{Pons:1988tj}
J.~Pons, {Ostrogradski Theorem for Higher Order Singular Lagrangians}, {\em
  Lett. Math. Phys.} {\bf 17}, p. 181  (1989).

\bibitem{Woodard:2015zca}
R.~P. Woodard, {Ostrogradsky's theorem on Hamiltonian instability}, {\em
  Scholarpedia} {\bf 10}, p. 32243  (2015).

\bibitem{deUrries:1998obu}
F.~de~Urries and J.~Julve, {Ostrogradski formalism for higher derivative scalar
  field theories}, {\em J. Phys. A} {\bf 31}, 6949  (1998).

\bibitem{Todorinov2020}
V.~Todorinov, {Relativistic Generalized Uncertainty Principle and Its
  Implications}  (2020).

\bibitem{Balek:2019nqk}
P.~Balek, {Charged-hadron suppression in Pb+Pb and Xe+Xe collisions measured
  with the ATLAS detector}, {\em Nucl. Phys. A} {\bf 982}, 571  (2019).

\end{thebibliography}

\end{document}